\documentstyle[12pt]{article}
\begin{document}
\newcommand{\m}{\: \:{\scriptstyle {}_{\sim}^{<}}\: \:}

\begin{center}
{\Large \bf Nonlinear phase shift without cascaded second-order
processes and third order nonlinearity}\\
\bigskip
V.P. Drachev, S.V. Perminov\\
Institute of Semiconductor Physics, Siberian Brunch of RAS,
pr.~Lavrent'eva, 13, Novosibirsk, Russia
\end{center}
\begin{abstract}
The new mechanism for obtaining a nonlinear phase shift has been
proposed and the schemes are described for its implementation. As it
is shown, the interference of two waves with intensity-dependent
amplitude ratio coming from the second harmonic generation should produce
the nonlinear phase shift. The sign and amount of nonlinear distortion of
a beam wavefront is dependent of the relative phase of the waves that is
introduced by the phase element. Calculated value of $n_2^{\rm eff}$
exceeds that connected with cascaded quadratic nonlinearity, at the same
conditions.
\end{abstract}


\bigskip

There are a lot of potential applications for materials which exhibit a
nonlinear phase shift (NPS) of the incident beam due to a local change of
the refractive index. For instance, the nonlinear phase shift is necessary
to obtain for the new generation of all-optical devices. Typically
nonlinear phase shift has been achieved with an intensity-dependent
refractive index coefficient $n_2$ ($n=n_0 + n_2I$) due to third-order
nonlinear susceptibility $\chi^{(3)}$. Currently another nonlinear
physical process that produces a nonlinear distortion in the phase of a
beam appears to be of great interest
\cite{Ostrovskiy67,Yablonovitch72,Stegeman93,Schiek93,Belashenkov89}. The
phase shift utilizes the cascading of two nonlinear second-order
processes: the up-conversion of a fundamental beam to a second harmonics
(SH) and its subsequent downconversion back to the fundamental. A number
of possible schemes of all-optical devices are discussed in Ref.
\cite{Assanto95}.

This letter theoretically predicts a new mechanism for nonlinear phase
shift. It is introduced when two waves with intensity-dependent
amplitude ratio interfere. We also suggest optical schemes that one can
use to obtain the nonlinear phase shift of both fundamental wave and
its second harmonics. The estimations show that the magnitude of the
nonlinear shift may be several times as large as the one occurring due to
cascaded processes under the same value of exciting intensity and
nonlinear medium length.

The essence of the proposed technique can be clarified by the following
simple considerations.

The phase $\Phi$ of the wave produced by interference of two
waves is given, as it is known, by the expression:
\begin{equation}
\label{1}
\sin\Phi=\frac{\sin\phi_1+\frac{a_2}{a_1}\sin\phi_2}{\sqrt{1+
\frac{a_2^2}{a_1^2}+2\frac{a_2}{a_1}\cos(\phi_1-\phi_2)}}.
\end{equation}
where $a_1$, $a_2$, $\phi_1$, $\phi_2$ are the corresponding amplitudes
and phases of the interfering waves. If we happen to produce a dependence
of the ratio $a_2/a_1$ upon intensity, obviously the phase $\Phi$ becomes
dependent of the intensity. To produce this intensity-dependent
ratio $a_2/a_1$, we will use second harmonics generation.

Further calculations are shown on the example of two schemes. The first
allows one to have NPS in the second-harmonics beam, and the other outputs
an NPS in the fundamental wave.

The first scheme (fig.~\ref{scheme1}a)
\begin{figure}\refstepcounter{figure}\label{scheme1}\end{figure}
is essentially, a dispersion interferometer that makes use of orthogonally
polarized second harmonic waves \cite{Drachev93OiS}. Both waves in the
dispersion interferometer propagate along the same path. Waves of
different frequencies correspond to the two arms of the interferometer,
and the relative phase is gained owing to the dispersion of the phase
element. In fig.~\ref{scheme1}a: 1,3 --- frequency doublers that use type
II phase matching (although an analogous scheme can be constructed for the
type I), 2 --- phase element (like an optical wedge with a compensator),
4 --- filter that cuts of the fundamental wave, 5 --- polarizer or
polarizing beam-splitter. The elements have the following orientation:
the frequency doublers are oriented orthogonally, the fundamental wave
is polarized at 45$^0$ angle with respect to the optical axes of the
crystals. The axis of the polarizer 5 is at the same angle as the
polarization of the fundamental wave.

The described interferometer has following notable features owing to
orthogonal polarization of the two second harmonic waves. First, the
second harmonics generated by the first doubler is not downconverted
inside the second crystal due to the large phase mismatch. Second, the
SH waves are generated in the first and the second crystals
independently of each other and interfere behind polarizer 5.  Let's
write the complex amplitudes of the two second harmonic waves as
$a^{(2\omega)}_1\exp{(i\phi_1)}$, $a^{(2\omega)}_2\exp{(i\phi_2)}$. The
relative phase $(\phi_2-\phi_1)$ can be adjusted with a phase element,
in our case, due to variable thickness of the optical wedge $\Delta l$:
$\Delta(\phi_2-\phi_1)=\frac{2\pi\Delta
l}{\lambda/2}\left[n(\omega)-n(2\omega)\right]$, where $n(\omega)$ and
$n(2\omega)$ are refractive indices of the wedge material at
frequencies $\omega$ and $2\omega$. Assume that the nonlinear phase
effects inside each crystal are absent or negligible, i.e. $\phi_1$ and
$\phi_2$ are independent of the intensity. This implies the following
conditions:
\begin{equation}
\label{2}
\Delta kl_{1,2}\ll 1;\,\,\,\,\, \frac{n_2Il_{1,2}}{\lambda}\ll 1,
\end{equation}
where $l_1$, $l_2$ are the lengths of the first and second crystals,
correspondly, $\Delta k$ --- wave-vector mismatch between the fundamental
and second harmonics, $I$ --- the incoming fundamental intensity. The
stated approximation essentially means that cascaded second-order processes
and third-order nonlinearity do not result in any tangible phase shifts.

Let's show that the phase of the resulting wave $\Phi$ at the exit from
the dispersion interferometer depends on intensity.
In the simplest case, when one can neglect the group velocity dispersion,
beam walk-off effect and use exact phase matching approximation $\Delta
k=0$, amplitudes $a^{(2\omega)}_1$ and $a^{(2\omega)}_2$ are given by
\cite{Blombergen62}:
\begin{eqnarray}
\label{5}
a^{(2\omega)}_1=I^{1/2}\mbox{th}\,
\left(l_14\pi\chi^{(2)}I^{1/2}\right), \\ \nonumber
a^{(2\omega)}_2=\left[I-{(a^{(2\omega)}_1)}^2\right]^{1/2}\mbox{th}\,
\left(l_24\pi\chi^{(2)}\left[I-{(a^{(2\omega)}_1)}^2\right]^{1/2}\right).
\end{eqnarray}
Expressions (\ref{5}) holds true while there is no initial second
harmonics in the incident light. In the second crystal this condition is
maintained owing to orthogonal orientation of the crystals. Behind
polarizer 5, the waves interfere having the amplitudes
$a^{(2\omega)}_1/\sqrt{2}$ and $a^{(2\omega)}_2/\sqrt{2}$. Thus, the
dependence of the amplitude ratio of the interfering waves can be
calculated with the expression:
\begin{equation}
\label{6}
\frac{a^{(2\omega)}_2}{a^{(2\omega)}_1}=\frac{(1-
\mbox{th}^2{\cal K}l_1I^{1/2})^{1/2}\mbox{th}
\left[{\cal K}l_2I^{1/2}(1-\mbox{th}^2{\cal K}l_1
I^{1/2})\right]}{\mbox{th}\,{\cal K}l_1I^{1/2}},
\end{equation}
where ${\cal K}=4\pi\chi^{(2)}$.

Let's analyze qualitatively Eqs. (\ref{1}), (\ref{6}). For
$I^{1/2}\m\frac{1}{{\cal K}l_1}$ we have th\,$\alpha\sim\alpha$ and,
hence, can simplify the formula:
\begin{equation}
\label{7}
\frac{a^{(2\omega)}_2}{a^{(2\omega)}_1}
\approx\frac{l_2}{l_1}\left(1-{\cal K}^2l_1^2I\right).
\end{equation}
If $l_2 \gg l_1$, then $a^{(2\omega)}_2/a^{(2\omega)}_1 \gg 1$ at low
intensity and falls down as the intensity grows. In accordance with
(\ref{1}), at low intensity $\Phi\approx\phi_2$, and at high intensity
phase $\Phi$ is led to phase $\phi_1$.

The dependence of nonlinear phase shift $\Delta\Phi_{NL}$ upon
dimensionless input intensity ${\cal K}^2l^2I$ (where $l=l_1+l_2)$ is
shown in fig.~\ref{scheme1}b, calculated according to (\ref{1}),
(\ref{6}). The maximal nonlinear phase shift possible to achieve is
limited to the relative phase $\phi_2-\phi_1$, that explains the
saturation of the curve in fig.~\ref{scheme1}b.

In fig.~\ref{f4},
\begin{figure}\refstepcounter{figure}\label{f4}\end{figure}
we see the dependence of $\Delta\Phi_{\rm NL}$ and $I_{2\omega}/I$
upon the relative phase $(\phi_2-\phi_1)$. The nonlinear phase shift
can be either positive (leading to beam self-defocusing), or negative
(self-focusing), depending on $(\phi_2-\phi_1)$. It is easily seen that
the maximum shift is produced at $\phi_2-\phi_1\approx\pi n$, however
the SH output in contrast to it is at minimum, despite of nearly 100\%
second-harmonic generation efficiency at the given exciting intensity. The
reason is that only a part of SH radiation passes through the polarizer,
namely, the component which has the nonlinear phase shift. Of course, the
phase of the fundamental wave in this scheme is not dependent of the
intensity.

To obtain a nonlinear phase shift of the fundamental wave we can suggest
as an example a scheme similar to a polarizing interferometer with the
use of frequency doublers. The mechanism of the nonlinear phase shift
here is analogous to that of the dispersion interferometer, namely, due
to the interference of the waves with intensity-dependent amplitude ratio.
The scheme diagram is shown in fig.~\ref{f5}a.
\begin{figure}\refstepcounter{figure}\label{f5}\end{figure} The
interferometer consists of frequency doubler 1 where type I phase matching
occurs, polarization-sensitive phase element 2, filter 3 which cuts of
SH, and polarizer 4 (or polarizing beam-splitter).

We are interested in phase of the fundamental wave at the exit
from the interferometer. Assume the incident wave is linearly polarized
and has intensity $I$. Inside the crystal there will be an ordinary wave
of intensity $I\cos^2\xi$ and extraordinary wave of intensity
$I\sin^2\xi$, where $\xi$ is the angle between the incident polarization
and the polarization of the ordinary wave. After the polarizer, which
is oriented at $45^0$ with respect to the ordinary wave, there will be
interference between a fraction of the extraordinary wave and the
corresponding fraction of the ordinary one. Their amplitudes we
denote as $a^\omega_1$ and $a^\omega_2$, correspondly. Provided
that there is type I phase matching, the generation of SH will decrease
the amplitude of only one of the fundamental waves. Let's assume, for
certainty, the ordinary wave is partially converted into the
extraordinary wave of SH, and the other fundamental wave (losses
neglected) does not vary in intensity.  Obviously, the amplitude ratio
of the fundamental waves will depend on the input intensity:
\begin{equation}
\label{8} \frac{a^\omega_2}{a^\omega_1}=
\left[1-\mbox{th}^2({\cal K}lI^{1/2}\cos\xi)
\right]^{1/2}\frac{|\cos\xi|}{|\sin\xi|}.
\end{equation}
Varying angle $\xi$, we can thus change ratio
$a^\omega_2/a^\omega_1$, at low intensity.

Expanding (\ref{1}) with the expression given by (\ref{8}) we will have
the dependence of the resulting wave phase $\Phi$ upon intensity.

The dependence of the nonlinear phase shift of fundamental wave on the
incident intensity is demonstrated in fig.~\ref{f5}b, as well as the
dependence of output intensity on the incident one.

In the scheme shown in fig.~\ref{f5}a, the nonlinear distortion of the
fundamental beam in the output of interferometer (focusing or defocusing)
also takes place, which can be controlled by variation of relative phase.
The latter is changed by a polarization-sensitive phase element, in this
case.

As well as cascaded processes, the described mechanism of ultrafast
nonlinearity produces a nonlinear phase shift of a wave without changing
refractive index of a medium. These mechanisms of nonlinearity have,
essentially, the similar nature. Indeed, NPS occurring at cascading of
quadratic processes actually arises from the interference of the
incoming fundamental wave and the wave downconverted with a shifted
phase from second harmonics. The difference is in the way the
interfering waves are produced.

As an analogue to the nonlinear coefficient $n_2$ for cubic nonlinearity,
in parallel with second-order cascaded processes, one can introduce the
effective refractive index $n_2^{\rm eff}=\Delta\Phi_{\rm NL}c/\omega lI$
for the "interference" mechanism (two-wave interference with
intensity-dependent amplitude ratio). We compare $n_2^{\rm eff}$ for
"interference" phase shift of fundamental wave at $\Delta kl=0$ to
$n_2^{\rm eff}$, related to cascaded second-order processes calculated for
optimal mismatch $|\Delta kl|\approx3$ \cite{Belashenkov89,DeSalvo92}.
The comparison at ${\cal K}^2l^2I \le
10$, that corresponds to $I\le 22$~GW/cm$^2$ for 1-mm-thick KTP
\cite{DeSalvo92}, shows that the discussed interference mechanism gives at
$\phi_2-\phi_1\ge 3\pi/4$ the value of $n_2^{\rm eff}$ about $2\div3$
times as large as, cascaded second-order processes do, while at
$\phi_2-\phi_1\approx\pi/2$ these mechanisms produce comparable nonlinear
phase shifts.

The proposed technique makes it possible to produce nonlinear phase shift
without cascaded second-order processes and cubic nonlinearity. The
cascaded processes are eliminated by exact phase matching $\Delta k=0$ and
mutually orthogonal orientation of the crystals. In real materials, $n_2$
for fast cubic nonlinearity is much smaller than $n_2^{\rm eff}$ for
second-order processes, for instance, in KTP $n_2\approx0.1n_2^{\rm eff}$
\cite{DeSalvo92}. Nevertheless, if we allow for inertial cubic
nonlinearity ($\tau\approx1$~ns) \cite{Drachev96} for nanosecond pulses
$n_2$ exceeds $n_2^{\rm eff}$. For experimental observation of the
isolated effect in such case one must use picosecond pulses to eliminate
the influence of slow nonlinearity.

We have demonstrated a new method for obtaining fast nonlinear shift. The
effect has a simple and intuitive interpretation, and is related to
interference of two waves with intensity-dependent amplitude ratio caused
by second harmonic generation. Sign and value of the nonlinear phase shift
is given by the relative phase of the waves introduced by phase element.
The effect of "interference" nonlinear phase shift has properties similar
to those of the nonlinear phase shift caused by cascaded second-order
processes and has several times higher $n_2^{\rm eff}$ then the latter.
Furthermore, this effect allows producing NPS both of the fundamental
and second-harmonic waves.

It worth noting the following features of described technique.
As we can see in fig.~\ref{f5}b, the output intensity weakly
depends on the input one within a rather wide region. This property
seems to be attractive for pulsations damping. The strong dependence of
$\Delta\Phi_{NL}$ upon the relative phase near the point
$\pi$ (for instance, in fig.~\ref{f4} $\Delta\Phi_{NL}$ changes from
$0.95\pi$ to $-0.95\pi$ as $\phi_2-\phi_1$ varies within
$\pi/25$) can be potentially useful for dispersion interferometry of
nonlinear optical media \cite{Drachev96,Danilova96-OiS}.

The present investigation has been carried out with support of the
Russian Foundation of Basic Research, grant 96-02-19331, and the
Governmental Program for the Leading Scientific Schools, grant
96-15-96642. V.P.~Drachev is also thankful to ISF for support in the
frame of ISSEP, grant A61-44.

\centerline {\bf Figure Captures}
\bigskip

	Fig.~\ref{scheme1}. (a) --- The scheme for obtaining nonlinear phase
shift in second-harmonic radiation. (b) --- The nonlinear phase shift
$\Delta\Phi_{NL}$ vs ${\cal K}^2l^2I$; $l_2/l_1=2$,
$\phi_2-\phi_1=3\pi/4$.

	Fig.~\ref{f4}. $\Delta\Phi_{NL}$ (solid line, right axis) and
SHG efficiency (dotted line, left axis) vs the relative phase. Here,
$l_2/l_1=2$, ${\cal K}^2l^2I=25$

	Fig.~\ref{f5}. (a) --- The scheme of the device to produce a
nonlinear phase shift of the fundamental wave. (b) --- The nonlinear
phase shift (solid line, right axis) and the output fundamental
intensity (dotted line, left axis) vs ${\cal K}^2l^2I$. Here,
$\xi=0.2$, $\phi_2-\phi_1=3\pi/4$.

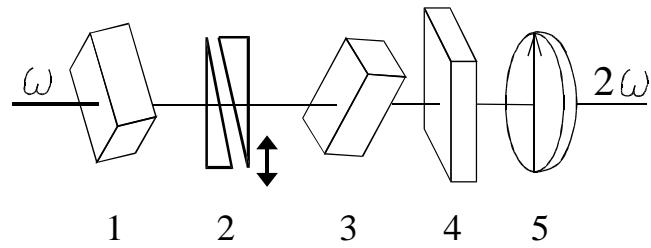
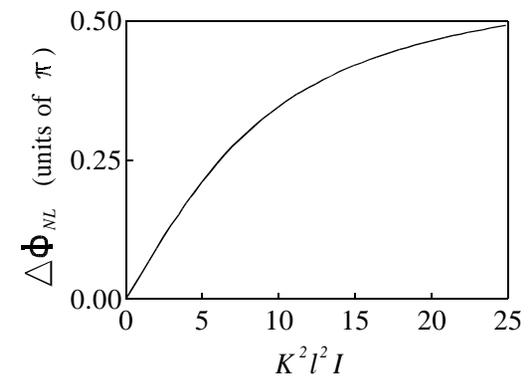
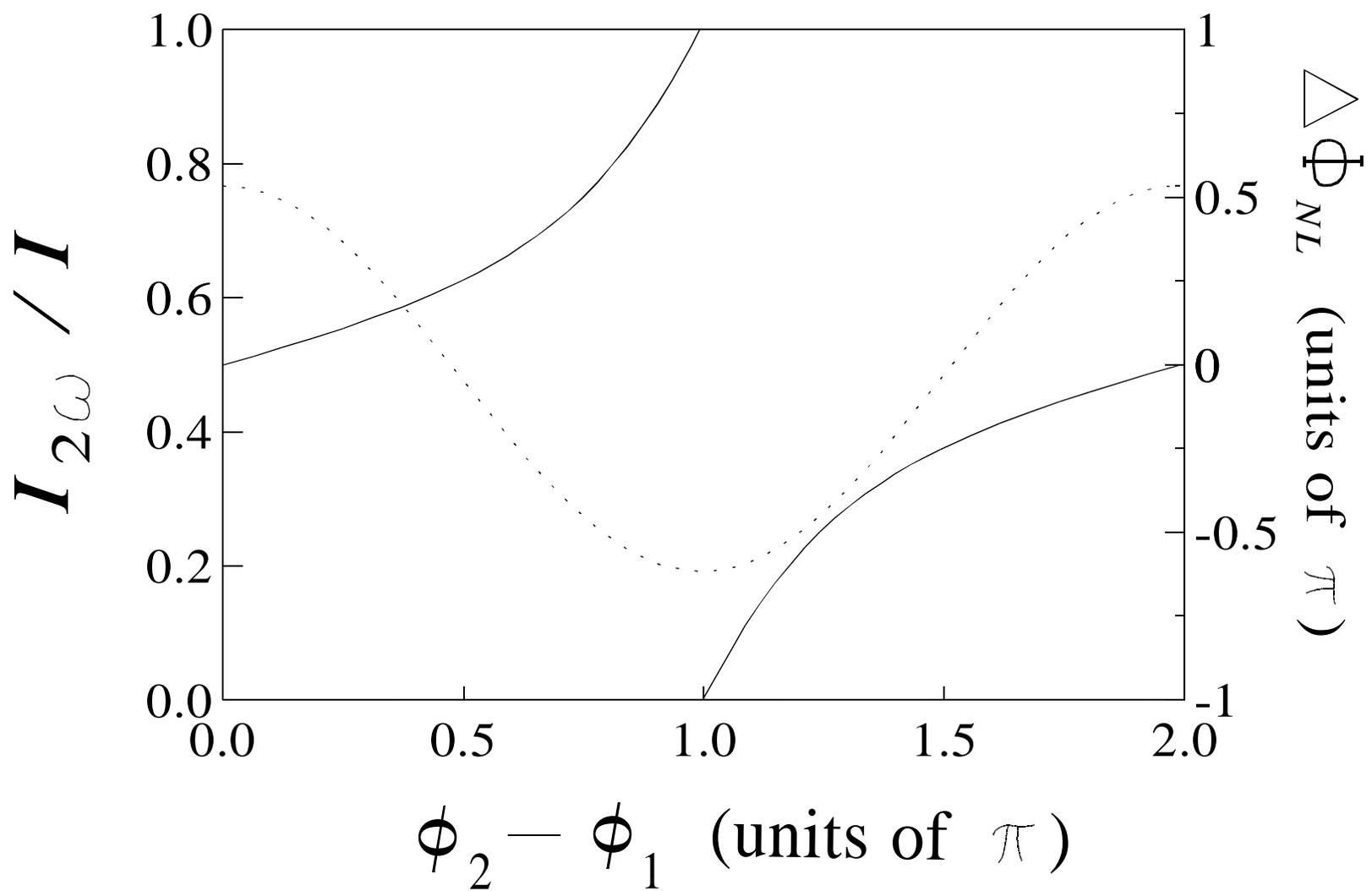
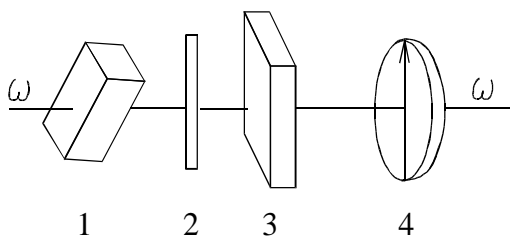
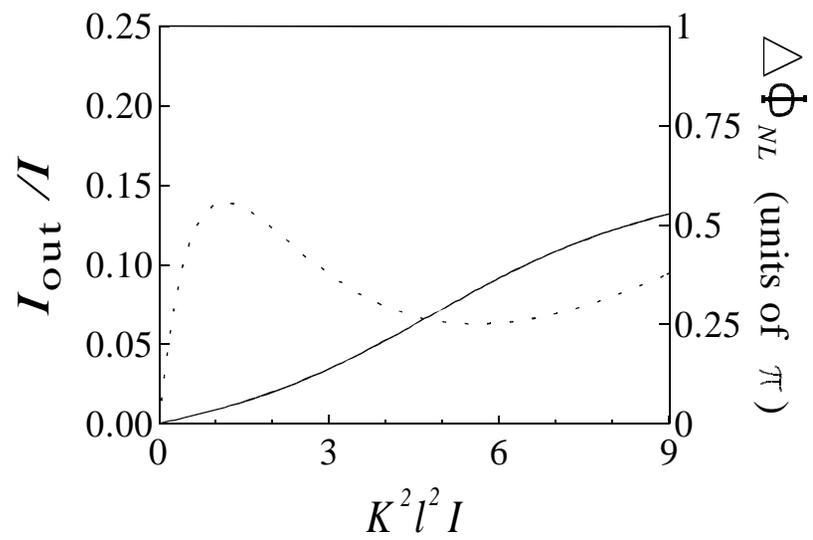
\end{document}